# A fractal climate response function can simulate global average temperature trends of the modern era and the past millennium

J.H. van Hateren

j.h.van.hateren at rug.nl
Institute for Mathematics and Computing Science, University of Groningen, The Netherlands

**Abstract** A climate response function is introduced that consists of six exponential (low-pass) filters with weights depending as a power law on their *e*-folding times. The response of this two-parameter function to the combined forcings of solar irradiance, greenhouse gases, and $SO_2$-related aerosols is fitted simultaneously to reconstructed temperatures of the past millennium, the response to solar cycles, the response to the 1991 Pinatubo volcanic eruption, and the modern 1850-2010 temperature trend. Assuming strong long-term modulation of solar irradiance, the quite adequate fit produces a climate response function with a millennium-scale response to doubled $CO_2$ concentration of 2.0 ± 0.3 ºC (mean ± standard error), of which about 50% is realized with *e*-folding times of 0.5 and 2 years, about 30% with *e*-folding times of 8 and 32 years, and about 20% with *e*-folding times of 128 and 512 years. The transient climate response (response after 70 years of 1% yearly rise of $CO_2$ concentration) is 1.5 ± 0.2 ºC. The temperature rise from 1820-1950 can be attributed for about 70% to increased solar irradiance, while the temperature changes after 1950 are almost completely produced by the interplay of anthropogenic greenhouse gases and aerosols. The $SO_2$-related forcing produces a small temperature drop in the years 1950-1970 and an inflection of the temperature curve around the year 2000. Fitting with a tenfold smaller modulation of solar irradiance produces a less adequate fit with millennium-scale and transient climate responses of 2.5 ± 0.4 ºC and 1.9 ± 0.3 ºC, respectively.

**1. Introduction**

In equilibrium, the earth receives as much energy from the sun as it radiates to space. An imbalance of incoming and outgoing radiation will change the temperature of the earth until a new equilibrium is reached. The temperature response to a step change in energy balance is called the climate response function. Its amplitude determines what the new equilibrium temperature will be and its shape determines how quickly that is reached.

The precise amplitude and shape of the climate response function of the earth is not known. It cannot be determined experimentally, and different models give different results depending on which processes are modelled in detail and on which values are assumed for the parameters involved (Randall et al 2007). A major uncertainty concerns the amount of heat transported between upper and deeper layers of the ocean. Such transport can significantly modify the shape of the climate response function.

In addition to uncertainty of model structure and parameters there is also uncertainty with respect to the energy flows driving the climate, the forcings. In particular, solar irradiance and the effects of anthropogenic aerosols are uncertain. Although solar irradiance has been well measured by satellites in recent decades, its variation is highly uncertain for the time before. Estimates of its variation over the past millennium vary by more than an order of magnitude (Wang et al 2005, Schrijver et al 2011, Shapiro et al 2011). In more recent times uncertainty is dominated by another forcing, the effect of anthropogenic aerosols as primarily produced by $SO_2$ emissions. Such aerosols have a direct cooling effect by reflecting incoming solar radiation to space, but also an indirect effect by changing the properties of clouds, affecting the earth's radiation balance in various ways (Forster et al 2007). The exact magnitude of these effects is hard to determine.

One way of dealing with the uncertainties in climate response function and forcings is to acknowledge them explicitly and utilize them as leeway for fitting model responses to observed temperatures. For the most elaborate climate models, this could at most take the form of some implicit tuning, because the multitude of parameters and long computation times make it impractical to exhaustively explore the parameter space and perform an explicit fit. On the other hand, for very simple climate models this approach is feasible and has already been performed with some success, for example by fitting to the temperature response to volcanic aerosol (Douglass and Knox 2005) and by studying the autocorrelation of temperature fluctuations (Schwartz 2007, 2008). However, these approaches assume that the climate response function has a very simple shape and is well described by a single *e*-folding time. They are therefore believed to produce inaccurate estimates of this function (Wigley et al 2005a, Foster et al 2008, Knutti et al 2008; see also Sect. 3.4).

In this article, I introduce a simple climate response function that lifts the restriction of a single *e*-folding time. It consists of a sum of exponentials covering a range of time scales (Li and Jarvis 2009, Friend 2011), but formulated in such a way that only a single parameter suffices to steer the balance between fast and slow components. I use this climate response function for fitting to measured temperature responses over a wide range of time scales, in particular the response to a volcanic eruption (time scale about a year), the response to solar cycles (about a decade), the temperature trend in the current century and a half, and the temperature changes in the past millennium. The fit provides constraints on the balance between fast and slow processes shaping the climate response function, and constraints on the balance between warming by greenhouse gases and cooling by aerosols in the current era. Furthermore, it shows how this balance is affected by assumptions on the extent to which solar radiation has varied over the past millennium. Finally, it produces estimates of the climate response to a doubling of the $CO_2$ concentration.

## 2. Data and methods

Temperature and forcing data were mostly obtained from public repositories, as detailed in Appendix 1. All computations in this study concern globally averaged temperatures. For the fits to temperatures in the past millennium, the reconstructions of Moberg et al (2005) and Mann et al (2009) are used. However, these reconstructions were made for the Northern Hemisphere (NH) because that is where most proxy data are found. From published modern temperature records, it can be readily observed that the most conspicuous difference between NH and global temperatures is that the former show larger modulations, both at short and long time scales. Presumably, this is related to the fact that the Southern Hemisphere contains a much larger percentage of oceans than the NH, and therefore damps fluctuations more strongly. I used the 1880-2010 NH and global temperature records of both HadCRUT3 and GISTEMP to estimate this demodulation quantitatively. This was done by subtracting the mean from the NH data, multiplying by a factor *m*, adding the mean again, and finally performing a fit to the global data with *m* and an additive offset as free parameters. This produces adequate fits with *m*=0.90 ± 0.02 for HadCRUT and *m*=0.82 ± 0.02 for GISTEMP. Assuming that a similar demodulation is a reasonable approximation also for the longer timescales of the Moberg and Mann data, I used *m*=0.86 to demodulate these data sets to obtain the estimate of global temperatures used in the fits below. As a control, I also performed fits using the NH data without demodulation, and found that this changed the parameter estimates only marginally (offset changed by 0.08, other parameters by less than 4%).

For the total solar irradiance (tsi), I used a recent reconstruction made by Shapiro et al (2011). In order to conform with recent estimates of the solar constant (Kopp and Lean 2011), the tsi was multiplied by 0.997 and the solar constant of 1360.8 was subtracted. The forcing corresponding to the tsi was obtained by multiplying by 0.7 (assuming an albedo of the earth of 0.3) and by 0.25 (the ratio of cross section and surface area of the earth). No correction for the fraction of UV in the tsi was made, because the influence of UV on surface temperatures is highly uncertain. For the calculations in Fig. 7c-d, the modulation of the sun's tsi was reduced tenfold by dividing deviations from the solar constant by ten. The amplitude of the solar cycles was kept approximately the same by isolating the cycles (by subtracting a trend from the original solar irradiance) and afterwards adding them with an appropriate weight to the demodulated irradiance.

SO$_2$ emission data are available from 1850-2005 (Smith et al 2011). Except for the solar forcing, I used the year 1800 as the starting year for computing model responses to forcings. I extended the SO$_2$ data to 1800-1849 by first noticing that the emissions very closely follow an exponentially growing curve from 1850-1900. I fitted an exponential to that part of the curve and used that to extend the curve backwards to 1800. The remaining SO$_2$ emission in the year 1800 is close enough to zero to be neglected as a discontinuity when starting the computation in 1800. All computations below are performed until 2010. I used two scenarios for the missing data on global SO$_2$ emissions between 2006 and 2010. For the first scenario, the SO$_2$ emission was held constant at the 2005 level. For the second scenario, the almost linear growth in SO$_2$ emissions observed between 2002 and 2005 was linearly extrapolated until 2010.

I used the AGGI (Annual Greenhouse Gas Index, NOAA) as a reference for the forcing produced by well-mixed greenhouse gases. However, the AGGI is only available from 1979. To extend this backwards to 1800 I used primary data sources on greenhouse gas concentrations following the procedure as detailed in Appendix 2.

Computations for this article were performed with the open-source R language (http://www.r-project.org/). All computations mentioned and all figures shown can be readily reproduced. The R-scripts I wrote can be obtained from http://bit.ly/u99X2d or upon request from the author. Most computations in this article involve first-order low-pass filters, which are filters governed by a first-order differential equation $\tau dy/dt + y = cx$, with $x$ input, $y$ output, $\tau$ the $e$-folding time, and $c$ a gain. This is commonly known as the filter describing the voltage response to current injected into an RC-circuit (a resistor in parallel with a capacitor). Such filters have pulse and step responses characterized by an exponential with an $e$-folding time (time constant, relaxation time) $\tau=RC$. For computing the response of a low-pass filter to an arbitrary input, the recursive computing scheme of van Hateren (2008) is used. See Appendix 3 for further details.

All error bounds given in this article are standard errors.

**3. Results**

3.1 A fractal climate response function

In this article, the earth is simplified to be an object characterized by a single average temperature. It is in thermal equilibrium when it receives, on average, as much energy from the sun as it radiates away towards space. When the energy balance is perturbed, the temperature changes until the resulting change in outgoing thermal radiation reestablishes equilibrium. For small perturbations, the system can be assumed to respond linearly in good approximation (see Sect. 4.3 for a discussion). The temperature response is then proportional to the forcing, i.e. the deviation from energy equilibrium.

This temperature response is not instantaneous, though, and its dynamics depends on the physics of the system. In its simplest form the physics can be represented by a heat capacity, storing heat and dominated by the upper layers of the oceans, and by a resistive element that quantifies how easily energy is radiated towards space. The electrical analogon of this system is shown at the left side of Fig. 1a as the resistor $R_1$ and the capacitor $C_1$. Both are connected to ground (zero) because all signals are defined relative to equilibrium, which is zero by definition. In physical reality, the ground of $R_1$ is space and the ground of $C_1$ might be a subsurface level in the oceans. Substituting temperature for voltage, the temperature response $\Delta T_1$ of such an RC-circuit to a step in forcing $\Delta I$ is well known: $\Delta T_1 = R_1 \Delta I (1 - \exp(-t/\tau))$ with $t$ the time and $\tau=RC$ the $e$-folding time. This response function is an example of a climate response function, defined here as the temperature response to a unit forcing step. Because forcings in climate science are generally given as power per unit earth surface, $\Delta I$ has dimension Wm$^{-2}$, and $R_1$ K/(Wm$^{-2}$). The equilibrium response to a unit forcing step, $\Delta T_1 = R_1 \cdot 1$, is called here the model's equilibrium sensitivity (when the response refers to a doubling of CO$_2$ rather than a unit step it will be explicitly stated).

Physical considerations strongly suggest that a single heat capacity, i.e. a single $e$-folding time, is too simple to represent the earth adequately. One complication is that only part of the water in the oceans can be directly heated up or cooled down. Only an upper layer of perhaps 50-100 m is mixed

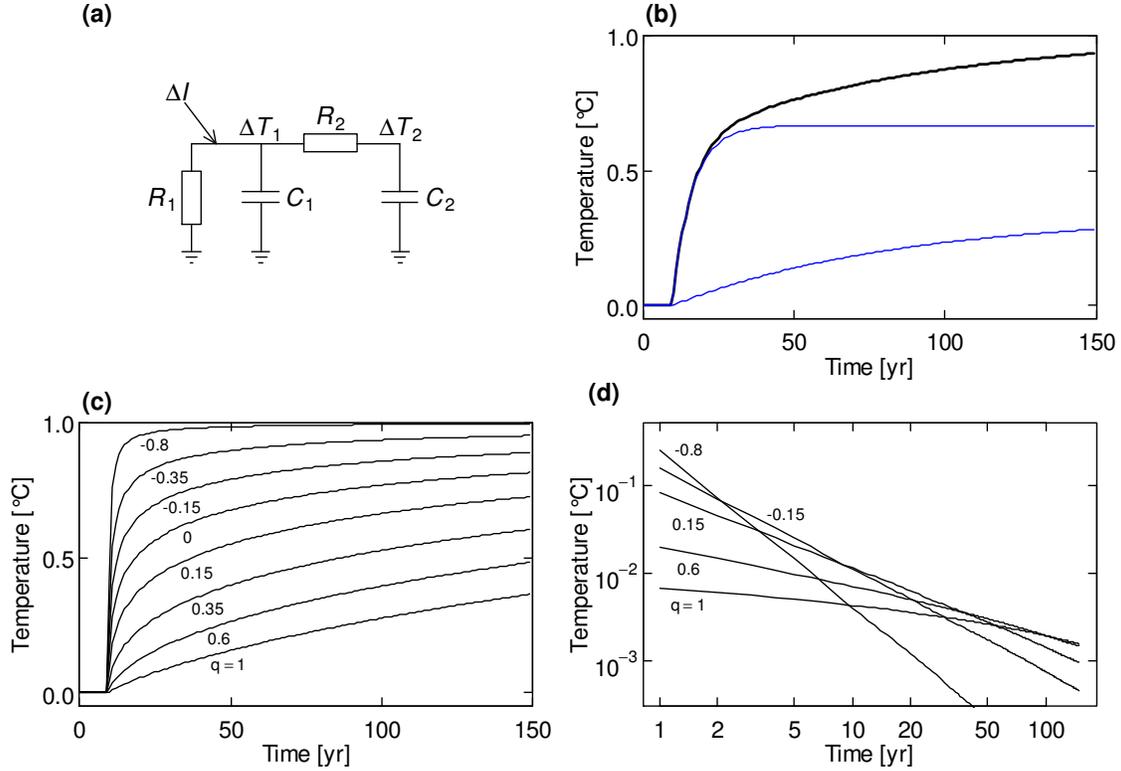

**Fig. 1 a** Equivalent circuit for a simple model of global temperature change $\Delta T_1$ in response to a forcing $\Delta I$. $C_1$ and $C_2$ represent the heat capacities of the ocean upper mixed layer and a deeper layer, respectively, $R_2$ the resistance to heat flowing between these two layers, and $R_1$ the resistance to heat radiating to space. See text for further explanation. **b** Black line: example response $\Delta T_1$ of the circuit in **a** to a step $\Delta I = 1$ Wm$^{-2}$, with $R_1=1$ K/(Wm$^{-2}$), $R_2=2$ K/(Wm$^{-2}$), $C_1=10$ Jm$^{-2}$/K, and $C_2=25$ Jm$^{-2}$/K. The response can be approximated by two exponential curves (blue lines). See text for further explanation. **c** Fractal climate response functions according to Eq. 4 with $n=6$. All curves go asymptotically to 1, but do so with different weightings of the six composing exponential curves (with $e$-folding times of 0.5, 2, 8, 32, 128, and 512 years). **d** Fractal climate pulse responses according to Eq. 1. The axes are both logarithmic, thus the approximately straight lines show power-law (fractal) scaling well enough by wind and other forces such that it can be considered a single heat capacity. Deeper layers can still exchange heat with the mixed upper layer, but only via mechanisms resistant to heat flow. The right-hand side of the circuit in Fig. 1a shows a simple model of this, where the heat capacity of deeper layers, $C_2$, is charged via a resistor $R_2$ connected to the mixed upper layer. Equilibrium is still defined as zero, thus in equilibrium the temperature deviations $\Delta T_1$ and $\Delta T_2$ are both zero.

The circuit of Fig. 1a was discussed in Schwartz (2008) and the corresponding equations analyzed in Held et al (2010). It can be shown that if $C_2$ is much larger than $C_1$ (assuming $R_1$ and $R_2$ are not very different) the response of this circuit is dominated by two $e$-folding times, a fast one $\tau_F = C_1 R_1 R_2/(R_1+R_2)$ and a slow one $\tau_S = C_2(R_1+R_2)$ (Held et al 2010). Fig. 1b shows an example with a numerical simulation of the step response (black line) and the two single exponential responses with $e$-folding times $\tau_F$ and $\tau_S$ (blue lines). The fast response dominates the step response until it reaches almost $R_2/(R_1+R_2)$ (2/3 in this example), where the slow response gradually takes over. When a new equilibrium is reached, $\Delta T_1$ must equal $\Delta T_2$, and all current $\Delta I$ flows through $R_1$. In other words, the amplitude of the equilibrium response is purely determined by $R_1$, not by $R_2$. Nevertheless, $R_2$ is important because it determines how much of the response is realized quickly and how much is realized slowly.

Although the circuit of Fig. 1a and its step response in Fig. 1b are likely more realistic than a single heat capacity, it is still a strong simplification of the physics. For example, the deeper layers of the ocean are insufficiently mixed to be represented by a single heat capacity $C_2$. Instead, a model with a range of layers with separate heat capacities and connected via separate resistances may be more appropriate. But a further complication arises because the ocean is not homogeneous across the globe, it has varying depth and the efficacy of heat exchange between upper and deeper layers is known to vary as well. A similar objection applies to $C_1$: shallow coastal waters are expected to produce a different local $C_1$ and therefore a different local $\tau_F$ than deeper waters. Moreover, about 30% of the earth's surface consists of land rather than water, again with different $\tau_F$. Concluding, it seems likely that a realistic climate response function consists of a continuous mix of components ranging from fast to slow. Large climate models that include ocean models with detailed physics indeed produce step responses that contain both fast and slow processes (see e.g. figure 1 in Friend 2011 and figures 3 and 6 in Hansen et al 2011).

As a parameterization of realistic climate response functions, I propose here a simple function that makes it possible to vary the relative weight of fast and slow processes without introducing an unwieldy number of free parameters. The function consists of a sum of first-order low-pass filters with their *e*-folding times and weights determined by a power law. Its pulse response is given by

$$h(t) = A_{\text{clim}} \sum_{i=1}^{n} \frac{a_i}{\tau_i} e^{-t/\tau_i}, \qquad (1)$$

with

$$\tau_i = r^{i-1} \tau_z \qquad (2)$$

$$a_i = \tau_i^q \Big/ \sum_{i=1}^{n} \tau_i^q, \qquad (3)$$

and its step response (i.e., the climate response function) by

$$R(t) = A_{\text{clim}} \left( 1 - \sum_{i=1}^{n} a_i e^{-t/\tau_i} \right), \qquad (4)$$

with Eqs. 1 and 4 only valid for $t \geq 0$; for $t < 0$ $h(t)=0$ and $R(t)=0$. As can be seen in Eq. 1, the pulse response consists of $n$ low-pass filters with *e*-folding times $\tau_z$, $r\tau_z$, $r^2\tau_z$, ... and weights $a_i$ depending on $\tau_i$ as a power law with power coefficient $q$. If $q$ is zero, all filters are equally strong, if $q$ is positive the slow filters with long $\tau$ get more weight relative to the fast filters, and if $q$ is negative fast filters get more weight. The normalization of $a_i$ in Eq. 3 is such that the summation in Eq. 1 integrates to 1, and $A_{\text{clim}}$ is therefore the model's equilibrium sensitivity. Although the function has five parameters, $A_{\text{clim}}$, $n$, $r$, $\tau_z$, and $q$, three of these will be fixed for the remainder of this study (apart from a few control runs) based on physical considerations.

The parameter $\tau_z$ determines the fastest component of the function, and therefore limits the fastest processes that can be described. The fastest measured process considered in the fits below is the global temperature response to the Pinatubo volcanic eruption in 1991. The response to this eruption is known to be reasonably well described by its aerosol production filtered with a low-pass filter with an *e*-folding time of 6-7 months (Douglass and Knox 2005). In order to enable capturing this response, $\tau_z$ is fixed to 0.5 year.

The ratio $r$ between subsequent $\tau$ is not crucial for the results. However, if $r$ is large, $h(t)$ undersamples scale-space (it does not sufficiently cover the range of *e*-folding times required), and its Fourier transform (the transfer function) is undulating, which is unrealistic for a physical climate response function. If $r$ is small, the number of filters and computational load increases accordingly. Below, $r=4$ is used as a reasonable compromise.

With fixed $\tau_z$ and $r$, the value of $n$, the number of filters, determines the largest *e*-folding time present. Modelling studies suggest that the *e*-folding times contributed by various parts of the ocean lie in a 100-1000 year range (Stouffer 2004). This range is mostly covered by fixing $n=6$, which gives filters with *e*-folding times of 0.5, 2, 8, 32, 128, and 512 years. The remaining two parameters of $h(t)$, $q$ and $A_{\text{clim}}$, are used as free parameters in the fits below.

Figure 1c gives examples of the climate response function $R(t)$ (shown for $A_{\text{clim}}=1$). As stated above, the parameter $q$ determines the balance between fast and slow processes. Being able to vary the balance between fast and slow processes is related to the approach of Hansen et al (2011), who use three differently balanced variants of the climate response function as a Green's function for computing temperature trends. In Hansen et al (2011) the climate response function (the system's step response in the language of linear systems analysis) is convolved with the time derivative of the total forcing (where the derivative effectively decomposes the forcing into little steps), whereas here the more conventional method is used to convolve the forcing directly with the system's pulse response (Eq. 1). Within the context of linear systems analysis, these methods are identical. The present work differs from the Hansen et al (2011) approach by using a single shape-related parameter ($q$) for explicit fitting, by not fixing the value of the climate sensitivity to a predetermined value, and by using a wider range of temperature data to test the model's dynamics and obtain its parameters.

The method of describing a climate response function with several components with different $e$-folding times, like is done here, has also been used by Friend (2011), who uses $e$-folding times of 2, 20, and 200 years to describe a GISS ModelE climate response function. Similarly, Li and Jarvis (2009) show that a HadCM3 model response can be fitted with three $e$-folding times (of 4.5, 140, and 1476 years). In both studies, the fitted sum-of-exponential functions are just used as fixed descriptions of an existing climate response function, and not varied in a fit as is done here.

Examples of the pulse response function $h(t)$ are shown in Fig. 1d on a double-logarithmic scale. The approximately straight lines show that the pulse response approximately follows a power law, i.e., $h(t) \propto 1/t^\alpha$, where $\alpha$ depends monotonically on $q$. Obviously, this power-law behaviour only occurs for a range of times, not for $t$ close to zero or $t$ very large. A consequence of a power-law pulse response is that short perturbations have short effects and long perturbations have long effects. The reason is that short perturbations cannot effectively engage the filters with long $e$-folding times, because such perturbations are essentially filtered out. This is very different from the response of a single first-order low-pass filter, which always gives responses characterized by the same time scale. For a power-law process, the time scale of the perturbation influences the time scale of the response. Because of the power-law scaling properties of the climate response function used here, it can suitably be called a fractal climate response function.

It should be noted that the power-law properties of $h(t)$ and $R(t)$ are chosen primarily for heuristic reasons. It is not known if the climate response function has such scaling properties, nor is there a compelling reason why it should have them, other than the empirical fact that such scaling is often encountered in nature when processes are active over a range of spatial or temporal scales. The advantage of the present definition is that it gives a fair amount of flexibility with only a few parameters.

Finally, it should be noted that the specific form of Eq. 4 does not allow a simple and straightforward identification of its components with specific physical components in the climate system (see also Li and Jarvis 2009 for a discussion). The response functions are just phenomenological models of the combined result of all the physical components and processes contributing to forced responses and feedbacks (see also Sect. 4.3). Long-term temperature variations attributable to internal dynamics of the climate system are not captured by the method (see also Sect. 4.2).

3.2 Isolating the response to solar cycles

One of the forcing-response pairs I will use in the fits below is the response to solar cycles, which are small, roughly sinusoidal modulations of the solar irradiance with a period of about 11 years and an amplitude of about 1 Wm$^{-2}$ (peak-to-peak, satellite measurements). These irradiance modulations are closely related to fluctuations in the number of sunspots that can be observed at the surface of the sun. The influence of solar cycles on the globally averaged temperature is small and masked by much larger temperature fluctuations attributable to other processes. Isolating this tiny signal is therefore a difficult problem. I used two different methods, which produce consistent results.

The first method takes an approach similar to that of Lean and Rind (2008, 2009) and Kopp and Lean (2011). An estimate of the average global temperature over a particular period is fitted with the main known sources of variance: a rising trend (either linear or based on estimated forcings),

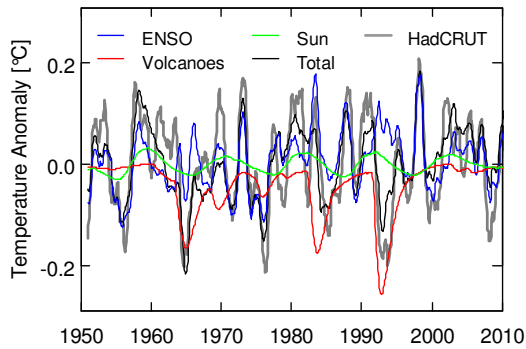

**Fig 2** Fit to the measured average global temperature (HadCRUT3) for the years 1951-2010, with forcings attributable to ENSO (El Niño Southern Oscillation), stratospheric aerosol originating from volcanic eruptions, and solar irradiance cycles. Temperature and solar irradiance were detrended with a loess polynomial fit before fitting, and ENSO, volcanic, and solar forcings were low-pass filtered as part of the fit

temperature fluctuations caused by El Niño and La Niña (El Niño Southern Oscillation, or ENSO below), temperature changes caused by volcanic eruptions, and the solar cycle modulations. A complication with this method is the following: the solar irradiance consists of cycles superimposed on a rising trend for much of the 1880-2010 period considered here. The temperature record from 1880-2010 shows a rising trend as well. A fit of the solar irradiance to the temperature record therefore attempts to match two things at the same time: the rising trend and the solar cycle modulations. Because the former is much larger than the latter, it is not clear how much of the resulting temperature cycles are truely present in the signal, or rather are artificially inflated because of fitting the trend at the same time. Adding a separate trend in a simultaneous fit does not fully solve this problem, because of the problem of collinearity (when fitting with correlated component signals the resulting weights are ambiguous) and because of uncertainty about the shape of the trend.

As a solution to the above problem, I removed the trends from both solar cycle and temperature records, before fitting. For this I used a loess polynomial fit (Cleveland et al 1992), with the span of the fit (0.3) chosen such that the base frequency of the solar cycle (1/11 cycle/year) was not significantly affected. Varying the loess span over a reasonable range produced values for the strength of the solar cycle response within the uncertainty range obtained below.

For the fits, I used the Multivariate ENSO Index (MEI; Wolter and Timlin 1998), available from 1950, and the extended MEI (Wolter and Timlin 2011) for the years before. Volcanic aerosol was according to data of Sato et al (1993; updated until 2010), and for the solar cycles the monthly sunspot number was used (the irradiance estimate by Wang et al 2005 gave similar results, but is not used here because it is only available until 2008). The ENSO, volcanic, and solar responses were each low-pass filtered during the fit with $e$-folding times $\tau_{ENSO}$, $\tau_{volc}$, and $\tau_{solar}$, respectively. The first two were used as free parameters in the fit, and gave values of 3-4 months for $\tau_{ENSO}$ and 4-12 months for $\tau_{volc}$. The fit diagnostics indicated that the solar cycle signal was too small to give reliable estimates of $\tau_{solar}$, and therefore $\tau_{solar}$ was fixed to 18 months (corresponding to the approximately 40º phase shift between solar cycle and response reported by White et al 1997).

It is known that the global temperature records contain a large artifact around the year 1945 (Thompson et al 2008) caused by an uncalibrated shift in the methods of temperature measurement by US and British ships. Following Zhou and Tung (2010), I therefore excluded the years 1942-1950 from the fits. To check for the sensitivity of the results to different epochs and different data sets, I performed fits to both HadCRUT and GISTEMP for 1880-1941, 1951-2010, and 1880-2010. An example of a fit is shown in Fig. 2. The amplitude of the solar cycle response was estimated by fitting a sinusoid to solar cycle 22 (using the years 1983-1998). For the example in the figure (HadCRUT 1951-2010) the fitted sinusoid had a peak-to-peak amplitude of 0.052 ± 0.007 ºC. Other results were for HadCRUT 1880-1941: 0.034 ± 0.012 ºC, 1880-2010: 0.057 ± 0.006 ºC, and for GISTEMP 1951-2010: 0.045 ± 0.007 ºC, 1880-1941: 0.000 ± 0.011 ºC, and 1880-2010: 0.039 ± 0.006 ºC. Clearly, the estimates vary considerably across periods and datasets, indicating a larger uncertainty than individual fits might let one to believe. Combining these estimates as a weighted mean, I find 0.043 ± 0.016 ºC. This is lower than the 0.08-0.10 ºC that is often mentioned in the literature (see Discussion).

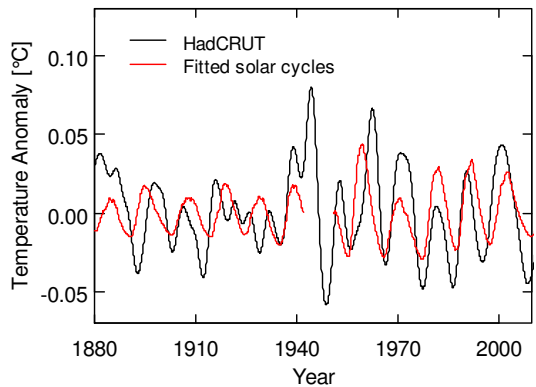

**Fig 3** Fitted solar cycle response (red line) to a processed HadCRUT global temperature record. The processing consisted of first splitting the record in 22-year segments, removing a linear trend and fitted estimates of ENSO and volcanic responses from each segment, and finally merging the segments. The resulting curve is shown after filtering by a gaussian ($\sigma$=18 months) for the purpose of presentation only (black line). The years 1942-1950 were excluded from the fit because they are known to contain a measuring artifact (see text for details)

I used a second method to isolate the solar response that does not depend on a loess trend removal, and that provides a residual solar signal to which the model response to solar cycles can be directly fitted. The method first divides the temperature record into segments of 22 years with 50% overlap between consecutive segments. The period of 22 years contains the response to approximately two solar cycles on average, and was chosen so that the removal of an offset and a linear trend is unlikely to affect the amplitude of the solar cycle response. After removal of a linear trend from each segment, a model consisting of an offset and responses to ENSO and volcanic aerosols was fitted to each segment separately. As before, the response to ENSO and volcanic aerosols were obtained through low-pass filters. Because the segments were too short to give reliable estimates of $\tau_{ENSO}$ and $\tau_{volc}$, these were fixed to 4 and 7 months, respectively. The model fits were subtracted, giving a residual response for each segment. Finally, these residuals were concatenated by using a complementary pair of $\sin^2$ and $\cos^2$ tapers to fade the signals in and out in the region of overlap of each pair of consecutive segments. This was done to avoid producing discontinuities at segment borders. The end result of this operation is a time series of the same length as the original global temperature series, but without trend and with ENSO and volcanic signals removed.

The residual time series was slightly low-pass filtered with a centred gaussian filter ($\sigma$=2 months) to reduce high-frequency noise that is outside the passband of the climate model. I fitted the response to solar cycles to this time series by using a low-pass filter with a fixed $\tau_{solar}$=18 months (see above). An example of such a fit is shown in Fig. 3, where the red curve is the fit to the raw residual time series. Because the raw data has too much variance to be useful to the human eye, it is shown in the figure as filtered with a gaussian with $\sigma$=18 months (black line). Note that the fit is fairly close for the period after 1950, apart from occasional phase mismatches, but that the solar signal appears to be only weakly present in the period before 1942. Fits for the same periods and datasets as before give the following amplitudes for solar cycle 22: HadCRUT 1880-1941: 0.038 ± 0.013 °C, 1951-2010: 0.056 ± 0.007 °C, 1880-2010: 0.053 ± 0.006 °C; GISTEMP 1880-1941: 0.002 ± 0.011 °C, 1951-2010: 0.047 ± 0.007 °C, 1880-2010: 0.037 ± 0.006 °C. Combining these estimates gives 0.044 ± 0.017 °C for the peak-to-peak amplitude of the solar cycle response. This is similar to the result obtained with the first method described above.

3.3 Isolating the response to the Pinatubo eruption

Large volcanic eruptions can inject aerosols into the stratosphere, which reflect part of the incoming solar radiation and therefore have a cooling effect on the globally averaged temperature. The effect is short-lasting, because the aerosols are fairly quickly removed from the stratosphere with an $e$-folding time of 0.8-1.5 year (Deshler 2008). The red line in Fig. 4 shows an estimate (Sato et al 1993) of the forcing (shown here in arbitrary units) produced by the 1991 eruption of the Pinatubo volcano. The black and blue lines show measurements of the effect on the global temperature. I isolated this signal from the HadCRUT and GISTEMP temperature records in the following way. First the period 1970-2010 was detrended with a loess fit (span=0.75). Second, a model consisting of an offset and responses to ENSO, volcanic aerosol, and solar cycles was fitted to the detrended temperature series. The response to the solar cycles was obtained with fixed parameters, such that it had the correct phase and an amplitude of 0.044 ºC as estimated above. Responses to ENSO and volcanic aerosol were obtained by filtering each with a low-pass filter; the free parameters in the fit were therefore offset, the *e*-folding times $\tau_{ENSO}$ and $\tau_{volc}$, and the two corresponding filter amplitudes. The part of the model consisting of offset, response to solar cycles, and response to ENSO was subtracted from the detrended time series, leaving a signal dominated by volcanic responses. The years 1989-2002 were selected as the Pinatubo period, and the response was defined relative to the mean 1989-1990 temperature.

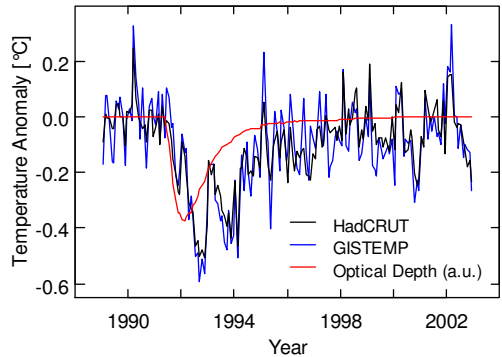

**Fig 4** The temperature response to the Pinatubo volcanic eruption in 1991, obtained by removing ENSO and solar responses from the temperature records of HadCRUT and GISTEMP. The red line shows (in arbitrary units) the estimated time course of the stratospheric optical density resulting from the volcano's aerosol (Sato et al 1993)

3.4 Fitting the model

The model fits are made with five free parameters. Two of these are $A_{clim}$ and $q$, defining the climate response function as discussed in Sect. 3.1. A third parameter concerns the forcing resulting from the Pinatubo eruption. The estimated time course (Sato et al 1993) of the change in stratospheric optical density (OD) that was caused by the aerosols Pinatubo produced is shown by the red curve in Fig. 4. The peak of this curve corresponds to an OD of approximately 0.15. The forcing caused by a unit OD is not exactly known. Wigley et al (2005b) cautiously suggests a value of -20 Wm$^{-2}$ per unit OD. Given the uncertainty, this factor is used as a free parameter, $A_{volc}$, and it is checked afterwards if the fitted value is reasonably close to the suggested value.

A fourth free parameter is an offset. Because all temperatures are anomalies, i.e., defined only relative to an arbitrary period, temperatures produced by a model can be offset without loss of generality. The offset also takes care of a degree of freedom in the solar forcing. The solar forcing is defined relative to the solar constant, taken as the total solar irradiance of 1360.8 Wm$^{-2}$ at the solar cycle minimum of 2008 (Kopp and Lean 2011). By having an offset as a free parameter, the (arbitrary) definition of the solar constant has no influence on the results.

Finally, the fifth free parameter is a factor, $A_{SO2}$, for converting the SO$_2$ emissions published by Smith et al (2011) into a forcing. This factor has a double role. First, leaving $A_{SO2}$ as a free parameter reflects the fact that the forcing produced by SO$_2$-related aerosols is only poorly known. The effects are partly local and therefore hard to quantify. Moreover, apart from direct effects through reflection of incoming solar radiation, there are also indirect effects on cloud formation and cloud longevity that are not fully understood (Forster et al 2007). In addition to taking care of this uncertainty, $A_{SO2}$ has a second important role. The other main anthropogenic forcing used in this study, the forcing attributable to well-mixed greenhouse gases, is determined fairly accurately (see Appendix 2). However, there are additional forcings, neither SO$_2$ nor well-mixed greenhouse gases, that need to be included. Examples are ozone and soot, both of which can act through several mechanisms that produce either positive or negative forcings (Forster et al 2007). The combined effect of these and other forcings is not well known. By leaving $A_{SO2}$ as a free parameter, this factor can absorb these other forcings. The assumption here is that the time course of the net effect of these other forcings has about the same shape as the SO$_2$-emission curve. This is most likely a strong simplification, but probably no more so than the other simplifications that are deliberately made in this article.

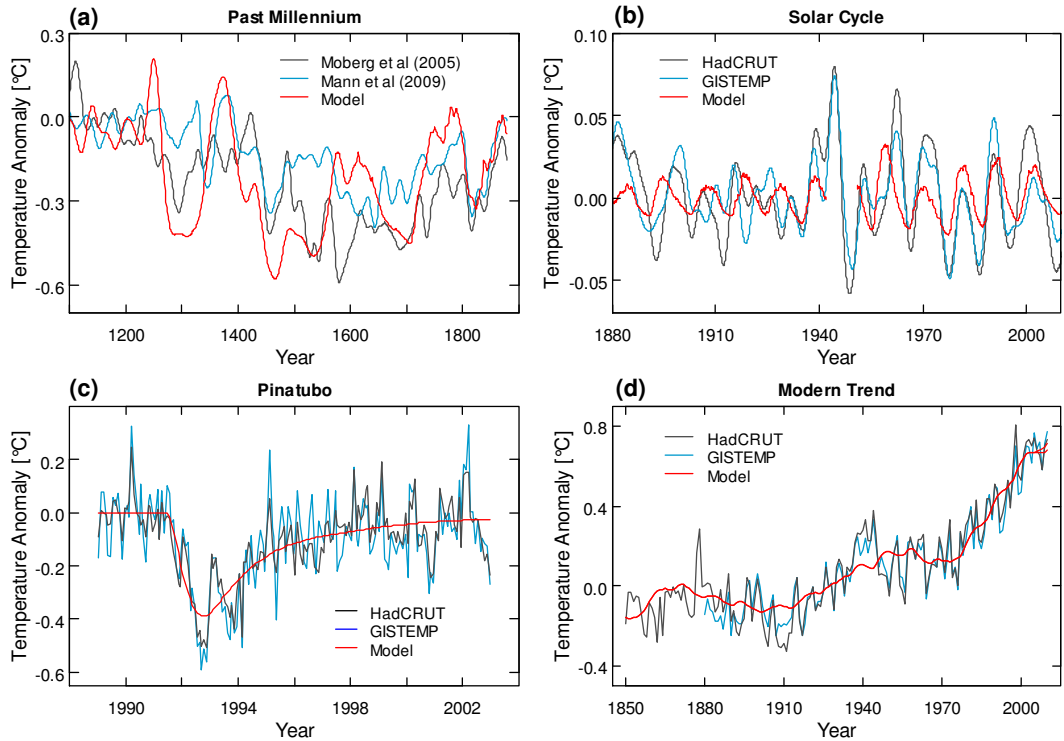

**Fig 5** The response of the model of Eq. 1 to forcings attributable to solar irradiance, greenhouse gases, and $SO_2$-related aerosol (see text and Section 2 for details), as fitted simultaneously to the four types of temperature records shown in **a**-**d**. In **a** the reconstructions of temperatures of the past millennium made by Moberg et al (2005) and Mann et al (2009) are shown, in **b** the response to solar cycles isolated from the HadCRUT and GISTEMP temperature records (as in Fig. 3), in **c** the response to the Pinatubo eruption (as in Fig. 4), and in **d** the modern temperature trends from HadCRUT (1850-2010) and GISTEMP (1880-2010). Anomalies of the measured temperatures in **a** and **d** are given relative to the mean temperature of 1880-1970

A fit was made to four different types of forcing-response pairs. Two of these were discussed above, namely the response to solar cycles (Sect. 3.2 and Fig. 3) and the response to the Pinatubo eruption (Sect. 3.3 and Fig. 4). Thirdly, the fit was made to reconstructions of the temperature of the past millennium. This was done for the years 1100-1880, using the reconstructions of Moberg et al (2005) and Mann et al (2009), which were slightly low-pass filtered with a gaussian with $\sigma$=5 years. Finally, the fit was made to the modern temperature trend, for the period 1850-2010 for HadCRUT3 and 1880-2010 for GISTEMP. Temperatures of past millennium and modern era are shown relative to the mean temperature of the years 1880-1970.

Figure 5 shows the result. It must be emphasized that the fit (red curves) is made simultaneously to all data shown. As can be seen, the overall fit to the four temperature records is quite adequate. Although all traces in Fig. 5a show considerable variance, they show clearly that both in measurements and model a somewhat colder period of approximately 1400-1850 (known as the Little Ice Age) is preceded and followed by generally higher temperatures. The fitted solar cycle response in Fig. 5b gives an amplitude of 0.038 ± 0.003 °C (peak-to-peak response to solar cycle 22), which is consistent with the estimate of 0.044 ± 0.017 °C discussed in Sect. 3.2. The phase shift between solar cycle forcing and response is 32°, consistent with the 30-50° phase shift reported by White et al (1997). The response to the Pinatubo eruption (Fig. 5c) is well fitted by the model, and gave a conversion factor of -15 ± 1 $Wm^{-2}$ per unit OD for obtaining forcing from aerosol optical density, which is of similar magnitude as the value of -20 $Wm^{-2}$ suggested by Wigley et al (2005b). Finally, also the modern trend is fitted quite well (Fig. 5d). Several of the temperature rises and drops that occur in the measured temperature series are present in the model response as well. A more detailed analysis of how the various forcings produce this model response is given in Sect. 3.6 below.

For the fit of Fig. 5d, only those forcings that contribute to a long-term trend are taken into account. In principle, forcings attributable to ENSO, volcanic eruptions, and other factors might be added. However, these forcings are expected to produce only fluctuations around a mean level, not a trend. For example, ENSO fluctuates up and down depending on the occurrence of El Niño or La Niña, and volcanic aerosol fluctuates up and down around a mean level of aerosol.

Apart from the Pinatubo forcing given above, the fit gave the following parameter estimates: offset $0.03 \pm 0.03$ °C, $A_{SO2}=(-8.8 \pm 1.0) \cdot 10^{-6}$ Wm$^{-2}$/(Gg SO$_2$), $q=-0.17 \pm 0.04$, and $A_{clim}=0.55 \pm 0.05$ K/(Wm$^{-2}$). It should be noted that the factor $A_{SO2}$ is not purely SO$_2$ related, but also absorbs some of the minor forcings (both positive and negative) that were not included in the well-mixed greenhouse gas forcing used here. The negative value of $q$ shows that the fast components in the climate response function are somewhat stronger than the slow ones (see the curve for $q=-0.15$ in Fig. 1c). About 50% of the response is realized in the short term (*e*-folding times 0.5 and 2 years), about 30% in the medium term (*e*-folding times 8 and 32 years), and about 20% in the long term (*e*-folding times 128 and 512 years).

As a test of the validity of the model with respect to the modern temperature trend, I performed fits to the data of Fig. 5 while excluding either the pre-1950 or the post-1950 modern temperature data from the fit. The fitted parameter values were subsequently used to compute the model response to the entire period. This led to fitted parameter values very similar to the ones given above (changes, in units as above, $\Delta$offset<0.02, $\Delta A_{SO2}<0.5 \cdot 10^{-6}$, $\Delta A_{volc}<0.01$, $\Delta q<0.002$, and $\Delta A_{clim}<0.01$) and model curves very close to the ones shown in Fig. 5.

As stated in the Introduction and in Sect. 3.1, there are physical arguments to assume that the climate response function is better described by multiple time scales than by a single exponential. As a test of this assumption, I implemented a model with a single *e*-folding time $\tau$ and fitted that to the data sets of Fig. 5. This produced adequate fits to the Pinatubo and solar cycle responses, a slightly reduced depth of the Little Ice Age minimum, and an underestimate of the 1970-2010 temperature rise. However, the fitted $\tau=1.2$ year is too small to be compatible with the ocean's thermal inertia. Fixing $\tau$ to a perhaps more realistic $\tau=5$ year (Held et al 2010) and fitting with the remaining four parameters produced adequate fits to millennial and modern temperature trends, but much too small and slow Pinatubo and solar cycle responses.

3.5 Climate sensitivities

In principle, the $A_{clim}=0.55 \pm 0.05$ K/(Wm$^{-2}$) found above gives the climate sensitivity and its uncertainty. However, the uncertainty is provided by the fit algorithm and it does not yet include the influence of parameters that were not varied in the fit. For example, varying the solar irradiance by 35% (Shapiro et al 2011 estimate a 20-50% uncertainty in their reconstruction) varies $A_{clim}$ from 0.47 to 0.63. Varying the effectiveness of the solar cycles by 40% (because of uncertainty in the role of UV in determining surface temperatures) varies $A_{clim}$ from 0.49 to 0.59. Several other parameters used in the analysis, in particular the ones that were used for obtaining the response to solar cycles, can also shift the value of $A_{clim}$ upwards or downwards. Because of these dependences, I believe the error bound of 0.05 in $A_{clim}$ obtained here from the fit is an underestimate. Assuming that an error of similar magnitude arises from uncertainty in parameters not included in the fit, I take as a fair estimate $A_{clim}=0.55 \pm 0.08$ K/(Wm$^{-2}$). With the forcing produced by doubling of CO$_2$, $3.7 \pm 0.3$ Wm$^{-2}$ (Gregory and Webb 2008) this gives a model response to a doubling of CO$_2$ concentration of $2.0 \pm 0.3$ °C. For the transient climate response (response to doubling produced by 70 years of 1% increase of CO$_2$ concentration per year, averaged over 20 years) I find $1.5 \pm 0.2$ °C, where the relative error is assumed to be similar to that of the model's equilibrium response.

The strength of the long-term component of the climate response function is quite uncertain, because the fit does not put a strong constraint on it. Varying the number of filters between $n=4$ and $n=7$ (equivalent to leaving out the *e*-folding times of 128 and 512, or adding one of 2048 years) only changes the quality of the fit marginally (slightly better for $n=6$ and $n=7$), but does vary $A_{clim}$ from 0.45 ($n=4$) through 0.50 ($n=5$) to 0.59 ($n=7$). Figure 6a shows the resulting response to a doubling of $CO_2$ concentration for $n$ ranging from 4 to 10 ($e$-folding times for $n=8$-$10$ are 8 ky, 33 ky, and 131 ky, respectively). Very slow components affect the response amplitude that is reached, but have negligible influence on the quality of the fits in Fig. 5 because the time scale of the data in Fig. 5 is limited to about 1000 years. I will therefore designate the response $2.0 \pm 0.3$ °C obtained above with fixed $n=6$ as the millennium-scale climate sensitivity, distinguishing it in that way from a fully equilibrated climate sensitivity that may contain components on time scales much longer than a millennium (see Sect. 4.3 for further discussion).

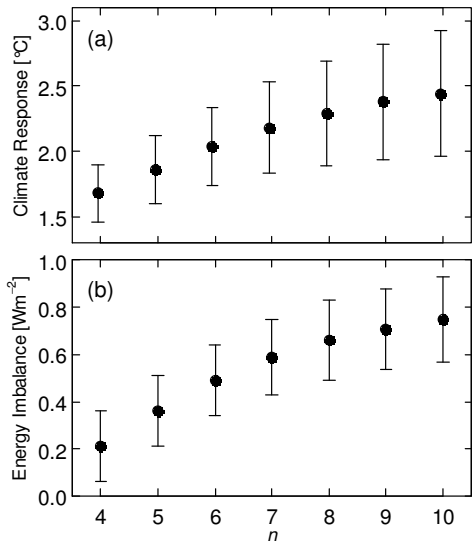

**Fig 6** Model properties as a function of the maximum time scale included in the response function. The model's equilibrium response to a doubling of $CO_2$ concentration is shown in **a**, and the computed earth's energy imbalance in **b**. The maximum time scales are given by the $n$ of Eq. 4, and correspond to $e$-folding times of 32 year ($n=4$), 128 y (5), 512 y (6), 2048 y (7), 8192 y (8), 32768 y (9), and 131072 y (10)

Adding components by increasing $n$ also changes the predicted power imbalance of the earth's climate system, because slow components add to the temperature rise that is not yet realized. From the forcing $F$, temperature $T$, and $A_{clim}$ it is possible to calculate this power imbalance as $F-T/A_{clim}$ (Hansen et al 2011), which is the power driving the global temperature in the direction of a new equilibrium. Figure 6b shows the power imbalance (commonly known as the energy imbalance) for the constant $SO_2$ scenario in the period 2001-2010. The estimates for $n=6$-$7$ are close to several recent estimates of the observed planetary energy imbalance: $0.59 \pm 0.15$ $Wm^{-2}$ (2005-2010; Hansen et al 2011) and $0.50 \pm 0.43$ $Wm^{-2}$ (2001-2010, 90% confidence range; Loeb et al 2012).

3.6 Checks and balances between different forcings

Figure 7 investigates in more detail how the fit to the modern temperature trend in Fig. 5d is obtained. Figure 7a shows the temperature contributions of solar irradiance, well-mixed greenhouse gases, and the factor that comprises $SO_2$-related aerosols and all other factors not separately included. The total temperature includes an offset (0.03 °C). Figure 7b shows the corresponding forcings, where the total includes an offset forcing of 0.06 $Wm^{-2}$. As can be seen, the temperature rise from about 1820 to 1950 is for the larger part (~70%) caused by increasing solar radiation, with the remainder caused by the net result of positive greenhouse gas forcing and negative $SO_2$ forcing. The fast rise in $SO_2$ emissions after about 1950 caused a slight drop in temperatures between 1950 and 1970, essentially because the subsequent rise in greenhouse gas forcing (primarily $CO_2$) lags the $SO_2$ forcing and is less steep.

From about 1970, the concerns in industrialized nations about the adverse effects of $SO_2$ emissions on human health and ecological vitality led to effective policies to reduce those emissions. The flat or even somewhat declining $SO_2$ curve and the rising $CO_2$ curve subsequently produced a steep increase in global temperatures between about 1970 and 2000. After the year 2000, $SO_2$

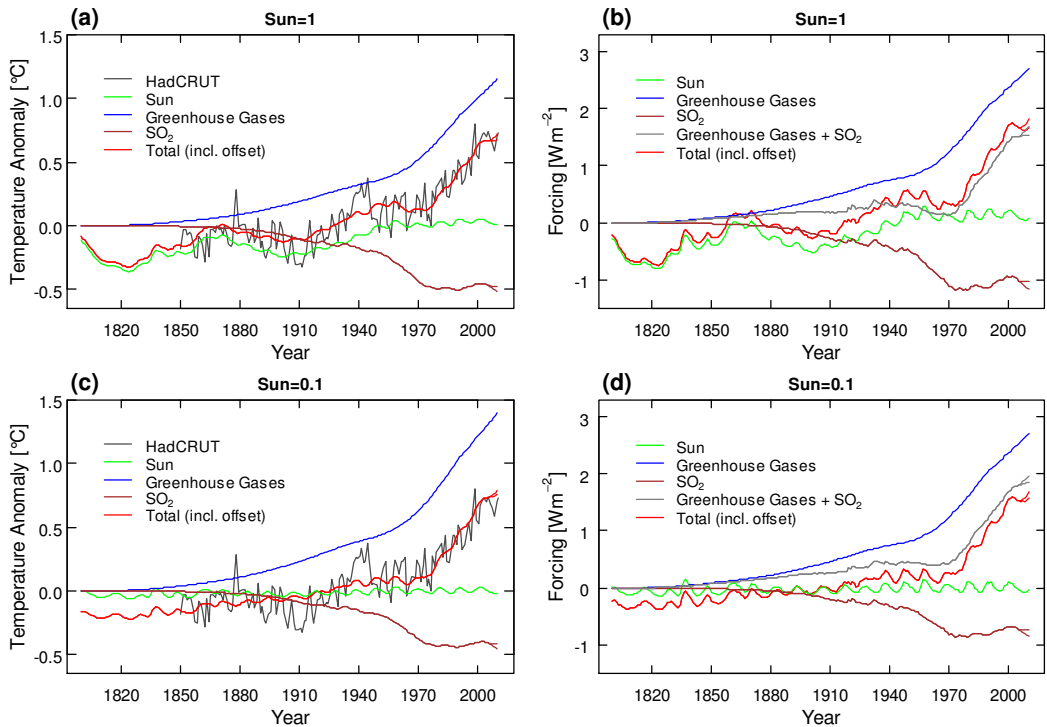

**Fig 7** Analysis of the contribution of the various forcings to the model as fitted to the modern temperature trend. The fit in **a** is identical to the one in Fig. 5d, thus made to HadCRUT and GISTEMP simultaneously, but only HadCRUT is shown here for the sake of clarity. The total temperature in **a** includes an offset of 0.03 °C. The forcings corresponding to the temperature responses in **a** are shown in **b**, where the total includes an offset of 0.06 Wm$^{-2}$. In **c** and **d** the results are shown when the fit is made with a solar irradiance with tenfold-reduced modulations (apart from the solar cycles, which are left intact). Totals include offsets of -0.16 °C and -0.24 Wm$^{-2}$

emissions started to rise once more, primarily attributable to increasing emissions in China (Smith et al 2011). The model shows an inflection of the temperature curve at about the year 2000. Whether the global SO$_2$ emissions continued growth after 2005 is not yet clear, and the calculation therefore shows two scenarios for 2006-2010. The first holds SO$_2$ emissions constant at the 2005 level, and the second lets them increase at the same rate as in 2002-2005.

The solar irradiance in the pre-satellite era is quite uncertain, and it is therefore interesting to see what the fit produces if we assume a much smaller modulation in solar irradiance than was reconstructed by Shapiro et al (2011). To this end, I reduced the amplitude of their reconstruction tenfold, while separately processing the solar cycles to keep these approximately the same. The resulting fit is fully adequate for the responses to the solar cycles and the Pinatubo eruption (not shown). The fit to the millennial temperature records is very poor because the modulation of the response is about eightfold smaller than in Fig. 5a (forcing tenfold smaller, but climate sensitivity 25% higher, see below). The fit to recent temperatures is shown in Fig. 7c-d. The rising trend in modern temperatures since the early 19$^{th}$ century is now produced with almost no involvement of the sun. Three major changes occur in the way the forcings are weighted. First, the forcing produced by SO$_2$ is reduced (compare Fig. 7b with d). Second, $A_{clim}$ becomes larger (compare the greenhouse gas curves in Fig. 7a and c). Third, the offset becomes larger and negative (-0.16 ± 0.02 °C rather than 0.03 °C as before). Although the overall trend is well fitted, the various rises and drops visible in the measured temperature record are not or less well captured. The parameters of the fit are, apart from the offset, $A_{volc}$=-13 ± 1 Wm$^{-2}$ per unit OD, $A_{SO2}$=(-6.4 ± 0.8) ·10$^{-6}$ Wm$^{-2}$/(Gg SO$_2$), $q$=-0.14 ± 0.04, and $A_{clim}$=0.68 ± 0.06 K/(Wm$^{-2}$). As before, we can assume a somewhat larger uncertainty of $A_{clim}$ than the fit provides, leading to a millennium-scale response to CO$_2$ doubling of 2.5 ± 0.4 °C, and a transient climate response of 1.9 ± 0.3 ºC. The dependence on *n* follows approximately a 1.25× scaled version of Fig. 6a, increasing from 2.0 ± 0.3 ºC at *n*=4 to 3.1 ± 0.7 ºC at *n*=10. The energy imbalance is nearly identical to the one shown in Fig. 6b.

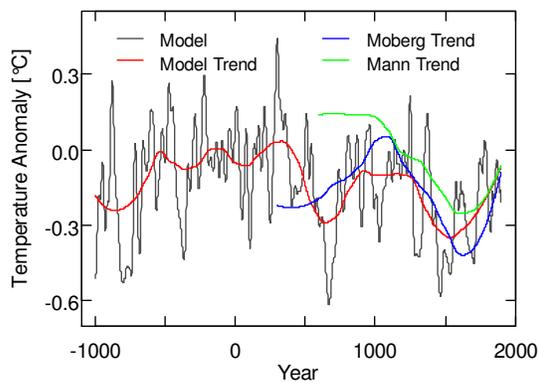

**Fig 8** Extrapolation of the model to past millennia. Because both the $^{10}$Be-based solar irradiance reconstruction and the proxy-based temperature reconstructions have large estimated errors, trends are shown as obtained from loess polynomial fits (with spans 0.25 for the model and 0.4 for the reconstructions of Moberg et al 2005 and Mann et al 2009). After about the year 1200 all three trends are fairly similar, but for earlier times this correspondence breaks down. The temperature produced by the model purely driven by reconstructed solar irradiance suggests warmer periods in Roman and medieval times, a fairly short colder period around the year 600 and a longer one at what is known as the Little Ice Age, about 1400-1850

3.7 Extrapolating to past millennia

The assumption for the fits of Figs. 5 and 7a-b is that variations in solar irradiance are the major cause of the Little Ice Age. It is interesting to see what the model predicts for even earlier times. Figure 8 shows this for the period -1000 – 1900 (grey line). The red line shows a trend obtained with a loess fit, and for comparison trends were obtained similarly for the temperature reconstructions of Moberg et al (2005) and Mann et al (2009). The three curves are fairly close together after the year 1200, essentially within the (large) error bands of the original temperature and irradiance data (see Fig. 5a for an impression of the variance). However, the Moberg and Mann reconstructions clearly deviate from the model response before the years 500 and 1000, respectively. This is not too surprising, because the temperature reconstructions are based on data, such as tree rings, that is quite scarce for early times.

The model trend suggests that, apart from colder periods around 500 – 700 and 1400 – 1850, there were warmer periods around -600 – 400 and 900 – 1200. These latter two periods are commonly known as the Roman Warm Period and the Medieval Warm Period. Recent work on South Pole temperature proxies (Bertler et al 2011) suggests that at least the Medieval Warm Period and the Little Ice Age were indeed global in nature.

3.8 Extrapolating to 2030

Predicting future temperatures requires, firstly, predicting future emissions of greenhouse gases and $SO_2$, as driven by global economic development, and, secondly, a detailed model of the carbon cycle, i.e., how for example $CO_2$ emissions lead to changes in $CO_2$ concentration (Eby et al 2009). Both requirements are beyond the scope of this article. However, for the very near future it is possible to make predictions based on simple extrapolations, which I will do here for the period until the year 2030.

The future forcing attributable to well-mixed greenhouse gases, in particular $CO_2$, is quite predictable on a decadal time scale. First, emissions are mostly generated by large installations with a long life time, such as power stations, and increases or decreases of emissions that are both large and abrupt are therefore unlikely. Second, $CO_2$ gradually accumulates in the atmosphere, which smoothes out fluctuations in emissions. Finally, forcing by $CO_2$ is proportional to the logarithm of the $CO_2$ concentration, which tends to linearize the accelerating $CO_2$ concentration curve expected from economic growth. For the years 2011-2030, I therefore linearly extrapolate the forcing as it occurred in 2001-2010.

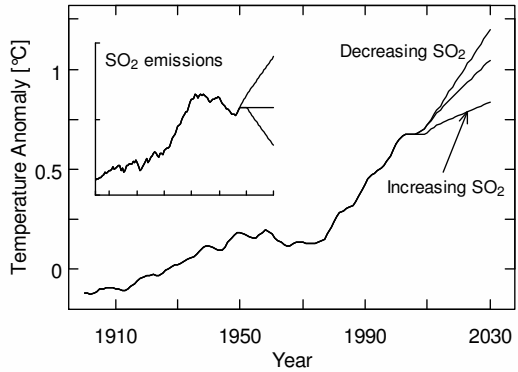

**Fig 9** Extrapolation of the model to the year 2030. Greenhouse gas forcing after 2010 is linearly extrapolated based on the 2001-2010 trend, solar forcing is assumed to remain constant at the mean 1950-2010 level, and $SO_2$ forcing is computed for three scenarios (see inset), either with decreasing $SO_2$ (2006-2010 at the 2005 level, followed by a yearly decrease of $2.5 \cdot 10^3$ Gg $SO_2$/year), constant $SO_2$ (2006-2030 remaining at the 2005 level), or increasing $SO_2$ (2006-2010 rising at the 2002-2005 rate, followed by a yearly increase of $2.5 \cdot 10^3$ Gg $SO_2$/year). The two axes of the inset encompass the years 1900-2030 and emissions of $0-2 \cdot 10^5$ Gg $SO_2$/year, respectively

Solar irradiance is assumed to remain at the mean level of 1950-2010, neglecting solar cycles after 2010. Although it is possible that the solar irradiance will start to change again like it presumably did in past centuries, such a change is not likely to be large in the coming 20 years (Solanki and Krivova 2011).

Much more uncertain are future $SO_2$ emissions. I compute three scenarios, one for constant $SO_2$ (2005 level all the way up to 2030), one for decreasing $SO_2$ (2005 level until 2010, followed by a yearly decline of $2.5 \cdot 10^3$ Gg $SO_2$/year, similar to the fastest decline of the Representative Concentration Pathways as shown in figure 7 of van Vuuren et al 2011), and one for increasing $SO_2$ (first linearly extrapolating the observed rise in the period 2002-2005 until 2010, followed by a yearly rise of $2.5 \cdot 10^3$ Gg $SO_2$/year). The inset of Fig. 9 shows these scenarios (the abscissa runs from 1900 to 2030, the ordinate from 0 to $2 \cdot 10^5$ Gg $SO_2$/year). The resulting temperature anomalies are also shown in Fig. 9. When $SO_2$ decreases quickly, the temperature will rise until 2030 with slightly higher speed than before 2000, whereas the rise with constant $SO_2$ is slightly slower because before 2000 $SO_2$ emissions were not constant, but slowly declining. For these scenarios, the 2000-2005 $SO_2$ increase remains only a small notch in the temperature record. Quite different is the result for the third scenario, with $SO_2$ emissions increasing. The global temperature then rises considerably less, because the increasing cooling effect of aerosols partly compensates for the increasing warming effect of greenhouse gases.

Note that curves as in Fig. 9 will be an excellent way to test the validity of the model developed here over the next one or two decades. As new $SO_2$ data along the lines of Smith et al (2011) becomes available, as well as new solar irradiance data from satellite measurements and updates of the AGGI, the model will predict global average temperatures, without free parameters. If the model is valid, these temperatures should match those from HadCRUT and GISTEMP.

**4. Discussion**

The results in this article show that a fractal climate response function, combined with assuming strong modulations of the solar irradiance, is consistent with temperature trends of past millennia as well as those of the modern era. In addition, the model is consistent with the temperature response to solar cycles and to the Pinatubo volcanic eruption. This is accomplished by a model fit that uses as its most important free parameters those parameters that are most uncertain in climate science: the climate sensitivity, the balance between fast and slow mechanisms of heat storage by the earth, and the effectiveness of cooling by aerosols of human origin.

4.1 Solar irradiance

A critical assumption in the approach taken here is that the recent reconstruction of solar irradiance in the pre-satellite era made by Shapiro et al (2011) is at least approximately correct. As discussed by Shapiro et al, the solar irradiance is uncertain, and indeed other recent reconstructions assume a ten-to twenty-fold smaller solar modulation (Wang et al 2005, Schrijver et al 2011). I investigated the consequences of a tenfold weaker solar modulation for the model fits in Fig. 7c-d, and found that the modern trend can still be fitted, although the fit appears less accurate than with a strongly modulated sun. A strongly modulated sun can explain the temperature modulation during the past millennium (Fig. 5a), as well as long-term temperature modulations that are believed to have occurred in earlier millennia (Fig. 8), whereas a weakly modulated sun can not. Explanations of the millennial trends by other forcings than the sun have been investigated, such as forcing produced by clustered volcanic eruptions, but the results remain somewhat unconvincing (Crowley 2000; Friend 2011). Moreover, the strength of volcanic forcing is not well known, because it is inferred from sulphate deposits that have an uncertain relationship with actual optical density. Optical density is strongly dependent on aerosol droplet size, which may have differed between eruptions (Timmreck et al 2009; see also Stothers 2007). The fits to the temperature trends in past millennium and modern era show that the present model has considerably more explanatory power with a strongly modulated sun than without one. However, only accurate, long-term observations of solar irradiance could provide decisive evidence on solar variability.

A recent article by Feulner (2011) shows that using the strong solar modulation of Shapiro et al (2011) as forcing for the CLIMBER-3α model produces results incompatible with temperature records, and argues that a strong solar modulation is therefore unlikely (see also Ammann et al 2007). This model has presumably been developed originally for use with a suit of forcings that includes the weakly modulated solar irradiance of Wang et al (2005), because that is the standard solar forcing used in almost all recent modelling efforts. It is then perhaps not too surprising if the model produces too large responses when it is driven by a strongly modulated sun, in particular as its 2×$CO_2$ transient and equilibrium responses, 2.3 and 3.6 ºC, are amongst the highest of similar models (Table 1 of Plattner et al 2008). There is apparently no solid evidence from solar physics that justifies preferring a weakly modulated sun over a strongly modulated one, it is an open question at this point in time (Shapiro et al 2011). It would therefore be interesting to know if the model used by Feulner (2011) can be adjusted, with realistic parameter settings, to accommodate a strongly modulated sun.

In contrast to long-term solar irradiance changes, the 11-year solar cycle is much better known because various satellites have measured it. The response of the globally averaged temperature to these irradiance modulations I obtain here, 0.044 ± 0.017 ºC peak-to-peak, is smaller than values of 0.08 – 0.1 ºC often mentioned in the literature. Some of these differences can be explained by the fact that local responses to solar cycles can be much larger, either positive or negative (Zhou and Tung 2010), presumably because of indirect influences on wind, clouds, and precipitation patterns. For example, White et al (1997) concentrate on the response measured in various ocean basins. In Tung et al (2008), those locations on earth that give a large response are given more weight in the average than other locations, and the value they find is thus not a global average. Zhou and Tung (2010) compute the response of global sea surface temperatures to a spatial weighting profile, subtracting negative responses from positive ones. This provides no direct information on an average response as would be produced by a spatially homogeneous weighting.

Lean and Rind (2008) perform a fit not unlike the one I performed for Fig. 2. Their figure 2 suggests that the solar cycle produces approximately a 0.1 ºC peak-to-peak response. However, the solar irradiance they use contains both a trend and solar cycles, which makes the solar cycle part of the response quite uncertain (see Sect. 3.2 for a discussion). Foster and Rahmstorf (2011) make a fit to the period 1979-2010 (see Lean and Rind 2009 and Kopp and Lean 2011 for similar results on about the same period), and find a solar cycle amplitude of approximately 0.08 ºC peak-to-peak response for surface temperatures. However, the lag between solar cycle irradiance and temperature response obtained from the fit, 1 month, is inconsistent with the 30-50º phase shift reported by White et al (1997) and difficult to reconcile with the ocean's thermal inertia. Using the software for the Foster and Rahmstorf (2011) analysis as made available by the lead author on his weblog (see Appendix 1) I could reproduce their results, and found that with more realistic lags of 12 and 18 months the solar cycle response is reduced by about 40%. I could confirm the phase sensitivity of solar cycle fits for the 1979-2010 period using the first method of Sect. 3.2. Reducing $\tau_{solar}$ from 18 months to 1 month changed the 1979-2010 estimate (mean of GISTEMP and HadCRUT) from 0.050 to 0.081 ºC. For the 1880-2010 estimate (compounded result as in Sect. 3.2), these two values of $\tau_{solar}$ yielded 0.043 and

0.037 ºC, respectively. The second method of Sect. 3.2 gave similar results. The particular phase sensitivity of the 1979-2010 estimates may be related to the fact that the 1979-2010 $SO_2$ aerosol forcing has considerable signal power in the frequency band of the solar cycles (see Fig. 7b) and therefore probably interferes with the cycle fit. Concluding, there appears to be no solid evidence in the literature that points to a globally averaged temperature response to solar cycles that is double the value of 0.044 ± 0.017 ºC I find here.

4.2 The modern temperature trend

Figure 5d shows that the modern temperature trend between 1850 and 2010 is fitted quite well by the model. Even many of the modulations around the trend are present, attributed to fluctuations of solar irradiance for the years before 1950 and for the years thereafter attributed to the interplay of opposing forcings by greenhouse gases and $SO_2$-related aerosols (Fig. 7a-b). It should be noted, though, that the fit is not perfect. For example, the temperature minimum around 1910 is less deep in the model than in the measurements, and the subsequent rise in temperature is less steep. It is possible that most or even all of this discrepancy is caused by the intrinsic errors of the reconstructed solar irradiance. Shapiro et al (2011) use two different $^{10}$Be datasets for their reconstructions, one from Greenland and one from the South Pole. The one not used here (the cyan curve in figure 2 of Shapiro et al 2011) has a much steeper rise between 1900 and 1930, and peaks earlier than the one used here. Obviously, this does not mean that more credibility can be given to one dataset or the other; it just means that the uncertainty in the reconstruction is large enough to accommodate at least part of the deviations between model and measurements. Similarly, some of the deviations may be caused by errors or biases in the measured temperature records, or by errors in the estimated greenhouse gas and aerosol forcings.

Another source of deviations between model and measurements is the fact that the model is a strongly simplified version of reality. It includes only what are judged to be first-order effects, and even those in a schematized way. Furthermore, the model is strictly linear (see Sect. 4.3 below). It is therefore likely that the model response would deviate from measured temperatures even if all forcings were known to high accuracy.

Finally, it is possible that some of the deviations are related neither to model deficiencies nor to inaccurately known forcings and temperatures, but are caused by fluctuations of the temperature generated by internal dynamics of the climate system. The size and temporal properties of such natural variations are not well known, but modelling studies (Katsman and van Oldenborgh 2011, Meehl et al 2011) suggest that they are not negligible. For example, it is possible that some of the decadal variations seen in the modern temperature trend are partly related to internal dynamics. Similarly, such dynamics may contribute to long-term temperature variations as observed on centennial and millennial time scales, perhaps strongly so in the case that the long-term solar modulation would eventually be shown to be weak as in the Wang et al (2005) reconstruction rather than strong as in the Shapiro et al (2011) reconstruction.

4.3 Climate response

The analysis in this article assumes that there is a linear relationship between net forcing and resulting temperature response, at least for the range of temperatures that have occurred over the past millennium up to now. To the extent that this assumption holds, the climate response function obtained from the fit implicitly incorporates all processes, including the various feedbacks, that contributed to the climate response over this period. However, there are additional processes that can change the shape and increase the amplitude of the climate response function, in particular processes acting on time scales longer than a millennium and processes that display nonlinear positive feedbacks, such as release of greenhouse gasses from methane hydrates. There are indeed paleoclimatological indications that the long-term climate response is larger than is found here. Excluding very slow processes like changes in ice sheet cover (which might double the temperature response, Hansen et al 2008, 2011), recent paleoclimatological estimates of the equilibrium climate response are 2.8 ± 0.9 ºC (Hansen et al 2008; here computed with 3.7 Wm$^{-2}$ as the 2×$CO_2$ forcing) and

2.3 (+0.3/-0.6) ºC (Schmittner et al 2011, 66% probability range). The difference between the results of these studies is mainly ascribed to different estimates of Last Glacial Maximum temperatures and different dust radiative forcing (Schmittner et al 2011).

The millennium-scale response to doubling of the $CO_2$ concentration found here, 2.0 ± 0.3 ºC, thus has presumably not yet reached full equilibrium, and can therefore only be cautiously compared with the equilibrium climate response of the 2007 IPCC report (Meehl et al 2007). It is at the lower end of the range considered likely (2-4.5 ºC), and lower than its best estimate (3 ºC). A first reason for this difference, as mentioned above, may be that the present estimate does not involve components beyond a millennial time scale (see also Sect. 3.5 and Fig. 6). A second reason may be the assumption of low solar variability generally made in model computations of the last decade, which tends to drive the climate sensitivity up (and the $CO_2$ doubling response to 2.5 ± 0.4 ºC) in the calculations I present in Fig. 7c-d. This higher sensitivity can then explain the temperature rise from the early 19$^{th}$ century until about 1950 without much involvement of the sun, if it is accompanied by a reduced weight of the aerosol forcing and a change in offset.

A final reason for the difference may be *a priori* uncertainty with respect to the longest time scales of the climate response function, in particular those related to the ocean's thermal physics. Forcing fits as in Fig. 5 with fixed $q=0$ (about 33% short-term response, 33% medium and 34% long) and $q=0.15$ (about 20% short-term, 30% medium, 50% long) produces $CO_2$ doubling responses of 2.7 ± 0.4 and 3.6 ± 0.6 ºC, respectively, and 2001-2010 energy imbalances of 0.72 ± 0.13 and 0.93 ± 0.10 Wm$^{-2}$. These values of $q$ produce model responses that do not match the data as well as those with the fitted $q$: the Little Ice Age gets a little too cold, the response to solar cycles a little too small, the tail of the Pinatubo response too fat, and the 1970-2010 temperature rise a little too large. Nevertheless, the model responses are still fairly close to the data, and would probably be considered satisfactory if the model parameters had been set in advance, based on empirical estimates and modelled physics, rather than obtained *a posteriori* in a fit as is done here.

4.4 Future scenarios

The fit of Fig. 7a suggests that $SO_2$ aerosols have an influence on global temperature that rivals that of greenhouse gases, in particular in determining fast changes in the overall trend. This may seem surprising, but in fact, it follows from differences between the dynamics of the forcings. Much of the $CO_2$ remains in the atmosphere for a long time (removed with a dominant *e*-folding time of about 100 years, Eby et al 2009), whereas $SO_2$ is typically removed from the troposphere within days (Liu et al 2005). When emissions start to rise because of economic growth, the cooling effect of $SO_2$ is felt earlier than the warming effect of $CO_2$, because the latter only gradually asserts its full effect by accumulating. The effect of $SO_2$ also tends to be stronger initially, because its forcing is approximately linearly related to emissions and concentrations, whereas forcing from $CO_2$ is proportional to the logarithm of its concentration. However, this tendency of $SO_2$ to have a stronger immediate effect than $CO_2$ becomes irrelevant once $SO_2$ emissions are specifically regulated, and thus do not covary anymore with $CO_2$ emissions. This happened in much of the industrialized world after about 1970, and thereby contributed to the fast rise in global temperatures between 1970 and 2000 (Fig. 7).

The strong short-term role of $SO_2$ emissions is also exemplified by the scenarios for the years 2011-2030 shown in Fig. 9. Depending on assumed $SO_2$ emissions, they produce quite different temperature trends. It is not clear which scenario, or indeed an intermediate or more extreme version, is most realistic. The Representative Concentration Pathways (van Vuuren et al 2011) as used for the upcoming IPCC fifth assessment all project decreasing $SO_2$, assuming stringent air pollution policies increasing proportionally to income. It remains to be seen if this will indeed be realized in the short term, because stringency and lag of policy implementations have historically varied strongly between regions (Smith et al 2005). There are reports that measures to stabilize or reduce $SO_2$ emissions are gradually becoming effective in China (Xu 2011), but not in India (Lu and Streets 2011). Many other regions in Asia, Africa and South America are showing considerable economic growth as well, and it seems possible that global $SO_2$ emissions will first rise or remain steady for some time before eventually declining. While reductions of $SO_2$ emissions are beneficial for human health and ecological vitality, such reductions contribute to global temperature rise. The masking of global warming by $SO_2$ aerosols was discussed before, and was dubbed a 'Faustian bargain' (Hansen and Lacis 1990, Hansen et al 2011).

Although the scenario with increasing $SO_2$ in Fig. 9 shows only a moderate temperature rise, it should be realized that this would be accomplished by a balancing act of two ever-increasing forcings. These forcings have different spatial profiles: $SO_2$ emissions are unevenly spread over the globe, whereas the $CO_2$ concentration is fairly uniform. That means that even if the globally averaged temperature does not change much, the spatial differences between the balancing forcings may still produce significant local increases and decreases of temperature, with associated changes in wind, cloud, and precipitation patterns. It is possible that some of the tropospheric aerosol leaks into the stratosphere (Randel et al 2010). Forcing by stratospheric aerosol has yet another spatial profile: it is more even than that of tropospheric aerosol, but still different from that of greenhouse gases. The reason is that it reduces the incoming solar radiation, i.e. unidirectional radiation impinging on a spherical surface. This is spatially different from the effect of greenhouse gases, which modulate outgoing thermal radiation that is basically omnidirectional.

Finally, with respect to the more distant future than 2030, it should be noted that temperature modulations caused by changes in $SO_2$ emissions are most likely limited to a maximum in the order of 0.5 ºC (Fig. 7a,c), through a continual balancing of the effects of economic growth and pollution control. Similarly, temperature modulations caused by a strongly modulated solar irradiance are also limited to a maximum in the order of 0.5 ºC (Fig. 8), and much less if solar irradiance is only weakly modulated. In contrast, a rising $CO_2$ concentration has the potential to cause an eventual temperature change of at least an order of magnitude larger.

## 5 Conclusion

In this article, a multi-scale climate response model was fitted to temperature records encompassing time scales ranging from a year to a millennium. On assumption of the correctness of a strongly modulated solar irradiance (Shapiro et al 2011) and by using recent data on $SO_2$ emissions (Smith et al 2011) the model provides tentative explanations for conspicuous trends in global average temperature from Middle Ages up to now (Figs. 5, 7a-b, and 8). The Medieval Warm Period and the subsequent Little Ice Age are primarily attributed to a decreased solar radiation in the latter period. The rise of the temperature from the early 19th century up to about 1950, including the fast 1910-1940 rise, is for about 70% attributed to an increase in solar radiation. The increasing warming by $CO_2$ up to 1950 is partly offset by increasing cooling by $SO_2$. The slightly cooling climate of 1950-1970 is attributed to $SO_2$ cooling overtaking $CO_2$ warming because of fast economic growth without much pollution control. The warming of 1970-2000 is attributed to increasing warming by $CO_2$ and decreasing cooling by $SO_2$ because of stringent air pollution policies. Finally, the post-2000 period with an apparent lull in temperature rise seems to replay the 1950-1970 events, with now China displaying fast economic growth with, initially, little pollution control.

**Acknowledgements** I thank dr A.I. Shapiro for making the solar irradiance data available, I thank the many maintainers of climate databases and datasets, in particular the KNMI Climate Explorer, the NOAA Paleoclimatology Program, and NASA GISS, for providing easy access to the data, and I thank H.R. de Boer for comments on the manuscript.

**Appendix 1: Data**

Data below was accessed November 21, 2011. Temperature data sets used in this study are the global and Northern Hemisphere (NH) HadCRUT3 (Brohan et al 2005) obtained from the KNMI Climate Explorer (van Oldenborgh et al 2008) at http://climexp.knmi.nl/data/ihadcrut3_gl.dat and http://climexp.knmi.nl/data/ihadcrut3_nh.dat; the global and NH GISTEMP (Hansen et al 2010) obtained from KNMI http://climexp.knmi.nl/data/igiss_al_gl_m.dat and http://climexp.knmi.nl/data/igiss_al_nh_m.dat; the Moberg data (Moberg et al 2005) obtained from the NOAA Paleoclimatology Program

([ftp://ftp.ncdc.noaa.gov/pub/data/paleo/contributions_by_author/moberg2005/nhtemp-moberg2005.txt](ftp://ftp.ncdc.noaa.gov/pub/data/paleo/contributions_by_author/moberg2005/nhtemp-moberg2005.txt)), and the Mann data (Mann et al 2009) from KNMI [http://climexp.knmi.nl/data/inh_mann.dat](http://climexp.knmi.nl/data/inh_mann.dat). Solar irradiance data are from Shapiro et al (2011), kindly provided by dr Shapiro; data from Lean (2000) and Wang et al (2005) were obtained from [ftp://strat50.met.fu-berlin.de/pub/outgoing/_matthes/CMIP5_solardata/TSI_WLS_mon_1882_2008.txt](ftp://strat50.met.fu-berlin.de/pub/outgoing/_matthes/CMIP5_solardata/TSI_WLS_mon_1882_2008.txt), the PMOD solar reconstruction (Fröhlich 2000) from KNMI [http://climexp.knmi.nl/data/itsi.dat](http://climexp.knmi.nl/data/itsi.dat), and sunspot numbers are from NOAA [ftp://ftp.ngdc.noaa.gov/STP/SOLAR_DATA/SUNSPOT_NUMBERS/INTERNATIONAL/monthly/MONTHLY](ftp://ftp.ngdc.noaa.gov/STP/SOLAR_DATA/SUNSPOT_NUMBERS/INTERNATIONAL/monthly/MONTHLY). The MEI (Multivariate ENSO Index; Wolter and Timlin 1998) and extended MEI (Wolter and Timlin 2011) were obtained from [http://www.esrl.noaa.gov/psd/enso/mei/table.html](http://www.esrl.noaa.gov/psd/enso/mei/table.html) and [http://www.esrl.noaa.gov/psd/enso/mei.ext/table.ext.html](http://www.esrl.noaa.gov/psd/enso/mei.ext/table.ext.html). The volcanic stratospheric optical thickness (Sato et al 1993) was obtained from [http://data.giss.nasa.gov/modelforce/strataer/tau_line.txt](http://data.giss.nasa.gov/modelforce/strataer/tau_line.txt). The $SO_2$ emissions (Smith et al 2011) were obtained from [http://ciera-air.org/sites/default/files/Total%20SO2.xls](http://ciera-air.org/sites/default/files/Total%20SO2.xls). Data on greenhouse gases were obtained from NOAA AGGI at [http://www.esrl.noaa.gov/gmd/aggi/AGGI_Table.csv](http://www.esrl.noaa.gov/gmd/aggi/AGGI_Table.csv), from NOAA Law Dome (Etheridge et al 1998; MacFarling Meure et al 2006) at [ftp://ftp.ncdc.noaa.gov/pub/data/paleo/icecore/antarctica/law/law2006.txt](ftp://ftp.ncdc.noaa.gov/pub/data/paleo/icecore/antarctica/law/law2006.txt), $CO_2$ data from NOAA Mauna Loa (Keeling et al 1976; Thoning et al 1989) at [ftp://ftp.cmdl.noaa.gov/ccg/co2/trends/co2_annmean_mlo.txt](ftp://ftp.cmdl.noaa.gov/ccg/co2/trends/co2_annmean_mlo.txt), $CH_4$ data from NOAA Mauna Loa (Dlugokencky et al 2005) from NOAA at [ftp://ftp.cmdl.noaa.gov/ccg/ch4/in-situ/mlo/ch4_mlo_surface-insitu_1_ccgg_month.txt](ftp://ftp.cmdl.noaa.gov/ccg/ch4/in-situ/mlo/ch4_mlo_surface-insitu_1_ccgg_month.txt), greenhouse gases used as forcings in GISS 2004 CGM (Hansen et al 1998; Hansen and Sato 2004) at [http://data.giss.nasa.gov/modelforce/ghgases/GHGs.1850-2000.txt](http://data.giss.nasa.gov/modelforce/ghgases/GHGs.1850-2000.txt), and $N_2O$ data from NOAA HATS (Montzka et al 2011) at [ftp://ftp.cmdl.noaa.gov/hats/n2o/combined/HATS_global_N2O.txt](ftp://ftp.cmdl.noaa.gov/hats/n2o/combined/HATS_global_N2O.txt). The software for the analysis of Foster and Rahmstorf (2011) discussed in Sect. 4.1 was obtained from [http://tamino.wordpress.com/2012/01/21/2011-temperature-roundup/](http://tamino.wordpress.com/2012/01/21/2011-temperature-roundup/) (accessed February 10, 2012).

**Appendix 2: Greenhouse gas forcing**

The AGGI (Annual Greenhouse Gas Index, NOAA) is only available from 1979. To extend this backwards to 1800 I used the following procedure. For $CO_2$ I obtained the Law Dome (South Pole) data and plotted that alongside the Keeling/NOAA Mauna Loa $CO_2$ measurements (1959-2010). Noticing a slight delay of the South Pole $CO_2$ concentration with respect to Mauna Loa I shifted the South Pole data 1.5 years forward, merged it with the Keeling/NOAA data, and fitted a loess curve (a local polynomial fit, Cleveland et al 1992) to the irregularly spaced result, using a span of 0.3 in the fit. The loess curve was subsequently used, with yearly spaced samples. In order to avoid discontinuities at the year 1800 I defined the 1800-1804 average $CO_2$ concentration as the reference, and tapered the $CO_2$ concentration gradually towards this value using a complementary $\sin^2$ and $\cos^2$ taper between 1800 and 1850. Forcing attributable to $CO_2$ with respect to the reference year 1800 was then calculated using the equation given in Table 6.2 of the Third Assessment Report of the IPCC (Ramaswamy et al 2001).

    For $CH_4$ I also used the Law Dome data, shifting it forward by 1.5 years and comparing it to $CH_4$ data collected at Mauna Loa (1988-2010). The Law Dome data was scaled up by 5.9% to match the Mauna Loa data, the data sets were merged, and a loess curve (span=0.3) was fitted for use with yearly spaced samples. As before, the 1800-1804 average was defined as the reference, and the data was similarly tapered towards 1800. For $N_2O$ the Law Dome data overlaps with $N_2O$ measured at Mauna Loa without further processing. The data was merged and fitted with a loess curve as before. With $N_2O$ clearly rising later than $CO_2$ and $CH_4$, the reference interval was taken as 1825-1829. After tapering as before, the forcings attributable to $CH_4$ and $N_2O$ were computed using the equations given by Ramaswamy et al (2001).

    After small corrections to the AGGI forcings for using slightly different reference values of pre-industrial $CO_2$, $CH_4$, and $N_2O$ than used above, the forcings determined above are nearly identical to those given by AGGI for these gases for 1979-2010. This means that for the period 1800-1978 the

forcings determined above are a consistent and probably quite accurate extension of the AGGI forcings. In addition to these three greenhouse gases, other compounds (CFC-11, CFC-12, and others) contribute to greenhouse forcing after about 1940. I used the concentrations of these compounds as used for GISS 2004 CGM computations to obtain forcings for CFC-11 and CFC-12 using again Table 6.2 of Ramaswamy et al (2001), and determining the effective forcing of the other compounds by matching the resulting total 1990-2000 forcing to the forcing given by AGGI. Finally the curves were smoothly connected by tapering over 1979-1989, with AGGI completely determining the curve from 1990 onward.

**Appendix 3: Computations**

For repeated computations, like when performing a fit, it is convenient to have a fast implementation that computes the response of a low-pass filter with *e*-folding time $\tau$ to an arbitrary input. I am using here a recursive computing scheme that is particularly fast and accurate (van Hateren 2008), specifically in the form of the Trapezoidal Rule (see Table 1 of van Hateren 2008) for $\tau$ sufficiently large compared with the sampling time $\Delta$ ($\tau/\Delta > 10$). The First-Order Hold (Table 1) was used for smaller $\tau$, because this scheme can also handle values of $\tau$ close to $\Delta$. The Modified Tustin's Method (Table 1) was used in a simple feedback configuration, following the methods outlined in van Hateren (2008), for numerical computation of the step response (Fig. 1b, black line) of the circuit of Fig. 1a. To avoid spin-up problems, the response to the solar irradiance was computed starting in the year -6000, with starting state given by the mean of the -6000 to 2010 solar irradiance.

A critical computation, filtering by the pulse response $h(t)$ of Eq. 1, was also implemented as a Fortran routine that was called from R. This gives identical results as the corresponding R routine, but reduces the computing time for the fit for Fig. 5 from a few minutes to a few seconds.

Most of the fits in this article were made with the nls algorithm for nonlinear least-squares fits that is part of the R language. This algorithm implements the nl2sol algorithm of the Port library, and provides error bounds for the parameters estimated in the fit. The model was simultaneously fitted to all data as shown in Fig. 5. Because the four different data types have different numbers of data points and different measurement errors, a weighted fit was performed with empirically determined weights, strongest for the response to solar cycles and the Pinatubo eruption, less for the modern temperature trend, and least for the temperature reconstructions of the last millennium. Weights were adjusted to be approximately in the middle of the range where visual inspection of the result showed that none of the datasets was basically ignored in the fit. Changing the weights by a factor of two up or down changed the offset by less than 0.02 and the other parameters by less than 12%.